\documentclass[lettersize,journal]{IEEEtran}
\usepackage{cite}
\usepackage{subcaption}

\usepackage{graphics} 
\usepackage{epsfig} 
\usepackage{booktabs}
\usepackage{colortbl}
\usepackage{multirow}
\definecolor{mygray}{gray}{0.9}
\definecolor{mydarkgray}{gray}{0.85}
\definecolor{darkgray}{RGB}{169,169,169}
\definecolor{darkgray176}{RGB}{176,176,176}
\definecolor{gray}{RGB}{128,128,128}
\definecolor{green01270}{RGB}{0,127,0}
\usepackage{svg}

\usepackage{moreverb,url}

\usepackage{hyperref}

\usepackage{blindtext}
\usepackage{hyperref}

\usepackage{times} 
\usepackage{amsmath} 
\usepackage{amssymb}  
\usepackage[capitalise]{cleveref}
\usepackage{algpseudocode}
\usepackage{algorithm}

\usepackage{stfloats}

\usepackage{lipsum} 

\newtheorem{definition}{Definition}[section]

\def\IR{{\mathbb R}}

\newtheorem{assumption}{Assumption}[section]
\usepackage{graphicx}
\usepackage{caption}
\begin{document}

\title{Obstacle Avoidance of Autonomous Vehicles: An LPVMPC with Scheduling Trust Region*}

\author{Maryam Nezami$^{1}$, Dimitrios S. Karachalios$^{1}$, Georg Schildbach$^{1}$ and Hossam S. Abbas$^{1}$
\thanks{* This work received partial funding from the German research foundation Deutsche Forschungsgemeinschaft
(DFG, German Research Foundation) – Projects No. 419290163. \\
$^{1}$ M. Nezami, D.S. Karachalios, G. Schildbach and H.S. Abbas are with the Institute for Electrical Engineering in Medicine, University of L{\"u}beck, L{\"u}beck, Germany (email: \{maryam.nezami, dimitrios.karachalios, georg.schildbach, h.abbas\}@uni-luebeck.de).\\ 
}
}

\markboth{Journal of \LaTeX\ Class Files, May~2024}%
{Shell \MakeLowercase{\textit{et al.}}: A Sample Article Using IEEEtran.cls for IEEE Journals}


\maketitle

\begin{abstract}
Reference tracking and obstacle avoidance rank among the foremost challenging aspects of autonomous driving.
This paper proposes control designs for solving reference tracking problems in autonomous driving tasks while considering static obstacles. 
We suggest a model predictive control (MPC) strategy that evades the computational burden of nonlinear nonconvex optimization methods after embedding the nonlinear model equivalently to a linear parameter-varying (LPV) formulation using the so-called scheduling parameter. 
This allows optimal and fast solutions of the underlying convex optimization scheme as a quadratic program (QP) at the expense of losing some performance due to the uncertainty of the future scheduling trajectory over the MPC horizon.
Also, to ensure that the modeling error due to the application of the scheduling parameter predictions does not become significant, we propose the concept of~\textbf{\textit{``scheduling trust region"}} by enforcing further soft constraints on the states and inputs. 
A consequence of using the new constraints in the MPC is that we construct a region in which the scheduling parameter updates in two consecutive time instants are trusted for computing the system matrices, and therefore, the feasibility of the MPC optimization problem is retained. 
We test the method in different scenarios and compare the results to standard LPVMPC as well as nonlinear MPC (NMPC) schemes. 
\end{abstract}


\section{Introduction}
Breakthroughs in technologies such as sensor technology and improvements in computing power have led to significant advancements in autonomous driving over the last decades. 
Model predictive control (MPC) is a control algorithm gaining popularity across various applications, including autonomous driving. 
A significant advantage of MPC lies in its direct enforcement of system and input constraints within the control algorithm. Consequently, MPC can generate optimal inputs that satisfy these constraints, making it suitable for safety-critical systems like vehicles. 

Nonlinear MPC (NMPC) is a commonly used MPC method that utilizes a nonlinear description of the system dynamics as the model of the system in the controller.  NMPC can make realistic estimations about the system's behavior by using a nonlinear model and/or nonlinear constraints. However, the accurate performance of NMPC comes with the cost of solving a nonlinear optimization problem online. 


Linear Parameter Varying (LPV) modeling is a method for representing nonlinear systems in a linear time-varying setting. It can be seen as a continuous smooth switching of adaptive linear models over nonlinear manifolds. In LPV models, the system matrices are functions of a vector called the scheduling parameter. The scheduling parameter itself is a function of states and inputs. The widespread application of MPC in different technologies has brought more attention to the employment of LPV models in MPC. 
An LPVMPC is a linear embedding of an NMPC algorithm that uses the structure of the problem and solves only a QP, similar to solving an optimization problem for an LTI system. However, one advantageous key difference between the LPV and LTI models is that in the LPV, the system matrices can be updated adaptively; thus, the LPV model better imitates the nonlinear system's behavior.


In order to build the future scheduling parameter vectors and system matrices when using an LPV model inside MPC, we require the future values of the states and inputs. However, since the future values of the states and inputs are decision variables of the optimization problem, they are not available. Therefore, we need to make predictions about the future scheduling parameter vectors. One common approach is to use the values of the scheduling parameter from the MPC solution at the 
previous time instant 
and use it in the current step. This approach has shown promising results under the condition that the scheduling parameter prediction error is not significant. However, in applications like obstacle avoidance while following a reference trajectory, the assumption of small scheduling parameter prediction error is not a valid assumption anymore because, in this kind of application, the system states and/or inputs have to change suddenly in order to keep the system safe. 
This issue can make the scheduling parameter prediction error relatively big 
and, therefore, reduce the applicability of LPVMPC. This is one of the problems that we tackle in this paper. 

\textit{Related works:} 
There is a large number of papers on the application of MPC in autonomous driving, e.g.,~\cite{kensbock2023scenario,zhang2020near,kim2021backup,shao2021model}, from longitudinal control for controlling the speed of the vehicle to lateral control for lane change and obstacle avoidance maneuvers. Also, with the recent rapid advancements in artificial intelligence-based control methods, there is growing attention to safe control architectures, which can provide safety guarantees for AI-based control methods. Because of the aforementioned advances of MPC, MPC is one of the common methods used in these safe control architectures to provide safety guarantees. e.g.,~\cite{brudigam2022safe, Nezami2021, li2020robust, wabersich2018linear,Nezami2022}.  

Using NMPC for problems like reference tracking or obstacle avoidance in autonomous driving is very common. In~\cite{allamaa2022real}, a practical NMPC framework for (real-time) obstacle avoidance with constant longitudinal speed is introduced. The suggested method assumes that a planner generates a new path when an obstacle is detected. Then, the NMPC follows the new reference to avoid the obstacle. However, by using this method, there is no guarantee that the vehicle will not hit the obstacle because the precise following of the new reference trajectory is not assured. In~\cite{arrigoni2022mpc}, a real-time trajectory planner based on NMPC is proposed. To ensure the NMPC is numerically solvable in real-time, an approach for solving the optimization problem by using a genetic algorithm is suggested. In~\cite{mi2023integration}, a nonlinear tube-based MPC for path planning and control is presented. In this approach, an extended kinematic vehicle model performs as the nonlinear vehicle model in the MPC. However, computing the tubes for the nonlinear model due to computational burden can be challenging. In \cite{incremona2022design}, an NMPC for autonomous driving is introduced, which takes care of driving on curvy roads, driving comfort, and pollutant emissions. It is shown that the NMPC is real-time applicable. However, for more complex driving scenarios, e.g., obstacle avoidance, more complex models that consider the tire slip angle should also be taken into account, which can increase the computation time significantly. 

Compared to NMPC, there are fewer papers on the application of LPVMPC for autonomous driving. 
In~\cite{rodonyi2021data}, it is shown that an LPV model for the steering dynamics of an autonomous vehicle could be identified from measured data. The paper also shows that the derived LPV model can deliver an efficient model of the dynamics even under extensive noise compared to the black-box identification of input-output models.
In~\cite{alcala2019lpv}, a control approach for autonomous racing by using LPV to model the dynamics of a vehicle and then using the LPV model in an LPVMPC framework for racing is suggested. In~\cite{alcala2020autonomous}, a cascade control method using an LPVMPC and an LPVLQR for autonomous driving is proposed. In~\cite{alcala2020lpv}, an LPV embedding of an NMPC for racing is introduced. Also, in this paper, the presence of obstacles in the race track is incorporated into the LPVMPC. 
In~\cite{tian2022gain}, an LPV model by taking the vehicle speed as the scheduling parameter and then, a robust path-tracking controller based on MPC and $\text{H}_{\infty}$ is presented. 
In~\cite{nezami2022robust}, a control architecture consisting of two subsequent controllers is proposed. At first, an MPC controls the longitudinal dynamics of a vehicle by providing values for the acceleration of the vehicle. Then, a robust LPVMPC controls the lateral dynamics of the vehicle to follow a reference trajectory. In this architecture, the longitudinal speed of the vehicle serves as the scheduling parameter in the lateral LPVMPC.

\textit{Contributions:} This paper 
presents a novel method for bounding the scheduling parameter prediction error in LPVMPC. In this approach, the concept of scheduling trust region is proposed which enhances the feasibility of the LPVMPC by keeping the change of scheduling parameter vectors in consecutive steps smooth.
The scheduling trust region is constructed online in the LPVMPC optimization problem using additional soft constraints on states and inputs of the system. 
Enforcing the scheduling trust region constraints in LPVMPC increases the LPVMPC range of applicability. 
The paper also compares the standard LPVMPC to LPVMPC with scheduling trust region and shows how the addition of the trust region constraint improves the feasibility of the optimization problem. 
Next, the LPVMPC with scheduling trust region and NMPC are evaluated against each other in terms of performance and computation time in several safe obstacle avoidance scenarios. 
The results are validated by using a high-fidelity vehicle model to represent the vehicle in the simulations.

\textit{Contents:}~\cref{sec:Preliminaries} presents some preliminary information about LPV modeling and its application in MPC. Then, in~\cref{Subsec:smoothschedul}, the concept of scheduling trust region for bounding the scheduling parameter prediction error is proposed. Next, in~\cref{Sec:Models}, at first, the nonlinear vehicle model, then the equivalent LPV model to the nonlinear vehicle model, is presented. Next, in~\cref{sec:NMPC}, the suggested NMPC followed by the method for generating the reference trajectory are introduced. In~\cref{sec:LPVMPC}, a method for linearizing the obstacle avoidance and road boundary constraints, as well as the novel LPVMPC setup, are presented. In the end,~\cref{sec:results} illustrates and compares the result of the application of the suggested LPVMPC to standard LPVMPC and NMPC designs.


\textit{Notations and definitions:} The notation $Q \succ 0$ represents the positive definiteness of a matrix $Q$. The weighted norm $\| x\|_Q$ is defined as $\| x\|_Q^2 = x^\top Q x $. The function $\text{diag}(\mathbf{x})$ constructs a diagonal matrix from a vector $\mathbf{x}$. A halfspace is defined as $\{ x \in \mathbb{R}^n | a^\top x \leq b\}$. The set of positive integers, including zero, is denoted by $\mathbb{Z}_{+} \cup \{ 0\}$. The symbol $\mathbb{I}_n$ denotes an $n$ by $n$ identity matrix where all elements on the diagonal are ones. The symbol $\otimes$ represents the Kronecker product. The symbol $\prod\limits_{i=1}^{N} A_i$ indicates product operation, i.e., $\prod\limits_{i=1}^{N}A_i = A_N A_{N-1} \ldots A_1$.

\section{Preliminaries}\label{sec:Preliminaries}

Consider the following representation of the LPV discrete-time system
\begin{equation}\label{eq:gen_lpv_model}
    z_{k+1} = A(p_k) z_k + B(p_k) u_k, \quad z_0 = \Bar{z}_0, \quad \forall k\! \in \!\mathbb{Z}_{+} \! \cup  \{ 0\},
\end{equation}
where $z_k \in \mathbb{R}^n$, $u_k \in \mathbb{R}^m$ and $p_k \in \mathbb{R}^{n_p}$, are the state vectors, the input vectors and the scheduling parameter vector of the system, respectively. The scheduling parameter vector can be a function of inputs and states, i.e., $p_k = g(z_k,u_k)$. Here, $\Bar{z}_0 \in \mathbb{R}^n$ represents the initial condition of the system.
The system~\eqref{eq:gen_lpv_model} is subject to state and input constraints as follows 
\begin{equation}~\label{eq:prel_input_state_cons}
\begin{aligned}
     &z_k \in \mathcal{Z} = \{ z_k \in \mathbb{R}^{n} | G^z z_k \leq h^z \}, \\
    &u_k \in \mathcal{U} = \{ u_k \in \mathbb{R}^{m} | G^u u_k \leq h^u \}, 
\end{aligned}
\end{equation}
where $\mathcal{Z}$ is the state constraint and $\mathcal{U}$ is the input constraint.

The LPV model~\eqref{eq:gen_lpv_model} can be used in an MPC as follows
\begin{subequations}\label{eq:gen_LPVMPC}
	\begin{align} 
		\underset{U_k,Z_k}{\text{min}} \
		& \| z_{N|k}   \|^2_P\! + \!\sum_{i=0}^{N-1}\left(\| z_{i|k} \|^2_Q +  \| u_{i|k}  \|^2_R\right)   \\
		 \text{s.t.} \;\;
		& z_{i+1|k} \!= \!A(p_{i|k}) z_{i|k} \!+\! B(p_{i|k}) u_{i|k},  \label{eq:gen_lpv_mpc_model} \\
		& z_{0|k} = \Bar{z}_0, \label{eq:gen_lpvmpc_init} \\
		&  z_{i+1|k} \in \mathcal{Z},  \label{eq:gen_lpvmpv_state_cons}\\
		&  u_{i|k} \in \mathcal{U}, \qquad \forall i = 0,1,\ldots,N-1,  \label{eq:gen_lpvmpc_input_cons}
	\end{align}
\end{subequations}
where the tuning matrices are $Q \succeq 0 \in \mathbb{R}^{n \times n}$, $R \succ 0 \in \mathbb{R}^{m \times m}$ and $P \succeq 0 \in \mathbb{R}^{n \times n}$. The horizon of MPC is denoted by $N$. The optimal sequence of inputs and states over the MPC horizon are $U^*_k = \{  u^*_{0|k}, u^*_{1|k}, \ldots, u^*_{N-1|k}  \}$ and $Z^*_k = \{  z^*_{1|k}, z^*_{2|k}, \ldots, z^*_{N|k}  \}$, respectively.

Analogous to the sequential approach of solving the optimization problem of an MPC for an LTI system, the above optimization problem can be rewritten as a quadratic program
\begin{subequations}\label{eq:gen_LPVMPC_seq}
	\begin{align} 
		\underset{U_k}{\text{min}} \
		& \frac{1}{2} U^\top_k \Bar{H}(P_k) U_k +  \Bar{F}(P_k)^\top U_k  \label{eq:gen_lpvmpc_cost} \\
		 \text{s.t.} \;\;
		& \Bar{G}^u U_k \leq \Bar{h}^u, \label{eq:gen_lpvmpc_input_cons_seq} \\
            & \Bar{G}^z \Gamma(P_k) U_k \leq \Bar{h}^z - \Bar{G}^z \Phi(P_k) \Bar{z}_0 \label{eq:gen_lpvmpc_state_cons_seq}, 
	\end{align}
\end{subequations}
where 
\begin{equation*}
    U_k = \begin{bmatrix}
        u_{0|k} \\
        u_{1|k} \\
        \vdots \\
        u_{N-1|k}
    \end{bmatrix}, P_k = \begin{bmatrix} 
        p_{0|k} \\ 
        p_{1|k} \\ 
        \vdots \\
        p_{N-1|k}
    \end{bmatrix}.
\end{equation*}
Also, $\Bar{G}^u = \mathbb{I}_{N} \otimes G^u$, $\Bar{h}^u = \mathbb{I}_{N} \otimes h^u$, $\Bar{G}^z = \mathbb{I}_{N} \otimes G^z$ and $\Bar{h}^z = \mathbb{I}_{N} \otimes h^z$, where $G^u$, $h^u$, $G^z$ and $h^z$ are from~\cref{eq:prel_input_state_cons}.
The estimations about the system behavior are made through the equation 
\begin{equation*}
    Z_k = \Phi(P_k) \Bar{z}_0 + \Gamma(P_k) U_k, 
\end{equation*}
where, $Z_k$, $\Phi(P_k)$ and $\Gamma(P_k)$ are presented below 
\begin{equation}\label{eq:z_psi}
    Z_k \!  =\!\!\! \begin{bmatrix}
       \! z_{1|k} \! \\
      \!  z_{2|k} \! \\
       \!  \vdots \! \\
       \!  z_{N|k} \!
    \end{bmatrix},  \Phi(P_k) \!= \!\! \!\begin{bmatrix}
        A(p_{0|k}) \\
        A(p_{1|k}) A(p_{0|k}) \\
        \vdots \\
       \prod\limits_{i=0}^{N-1} A(p_{i|k})  
    \end{bmatrix}, 
\end{equation}
\begin{equation}\label{eq:gamma}
    \Gamma(\! P_k\!) \! \!= \!\!\! \begin{bmatrix}
        B(p_{0|k}) \!\!& \!\!0 \!\!& \!\! \ldots & \!\! 0 \!\!\\
        A(p_{1|k}) B(p_{0|k})\!\! &\!\! B(p_{1|k}) \!\! & \!\! \ldots \!\! &\!\! 0 \!\!\\
        \vdots \!\! &  \!\! \vdots\!\! & \!\!\ddots \!\! &\!\! \vdots \!\!\\
       \! \prod\limits_{i=1}^{N\!-\!1} \!\!\! A(p_{i|k}\!)  B(p_{0|k}\!)  \!\! & \! \!\prod\limits^{N\!-\!1}_{i=2} \!\!\! A(p_{i|k}\!)  B(p_{1|k}\!)\!\! \! &\!\! \! \ldots \!\! \! & \!\! \! B(p_{N\!-\!1|k}\!)\!\!\!
    \end{bmatrix}\!\!.
\end{equation}
Then, the state constraint~\eqref{eq:gen_lpvmpv_state_cons}, turns into the constraint~\eqref{eq:gen_lpvmpc_state_cons_seq}.
Also, the matrices in the cost function~\eqref{eq:gen_lpvmpc_cost} are 
\begin{align*}
    \Bar{H}(P_k) &= 2(\hat{R} + \Gamma(P_k)^\top \hat{Q} \Gamma(P_k)), \\
    \Bar{F}(P_k) &= 2\Gamma(P_k)^\top \hat{Q} \Phi(P_k) \Bar{z}_0, 
\end{align*}
where, $\hat{Q} = \mathbb{I}_{N} \otimes Q$ and $\hat{R} = \mathbb{I}_{N} \otimes R$, with $Q$ and $R$ being from the cost of the original optimization problem~\eqref{eq:gen_LPVMPC}.

\section{Scheduling Trust Region} \label{Subsec:smoothschedul}
As can be seen in~\cref{eq:z_psi,eq:gamma}, the MPC formulation~\eqref{eq:gen_LPVMPC_seq} depends on the values of the future scheduling parameter $p_{i|k}$.
However, the exact value of $p_{i|k}$ is not always available. One practical approach, e.g., used in~\cite{abbas2024linear,nezami2022robust}, to handle this issue is to use the value of the scheduling parameter from the MPC solution at the previous time instant, as follows
\begin{equation}\label{eq:sch_pred}
    \hat{p}_{i|k} = p^*_{i+1|k-1}, 
\end{equation}
where $\hat{p}_{i|k}$ is the predicted scheduling parameter for step $i$ at time $k$ and $p^*_{i+1|k-1}$ is the value of scheduling parameter at step $i+1$ which was calculated at time $k-1$. 
Despite its imperfection, this method has proved to be effective in applications where the states and inputs change relatively smoothly, e.g.,~\cite{satzger2017robust, alcala2020autonomous}. 
However, assuming a constant scheduling parameter across consecutive times becomes an invalid assumption when the system states and inputs have significant changes in consecutive steps, making a challenge in the applicability of LPVMPC. This problem, however, has not received enough attention in the literature. In this section, we propose a practical solution to deal with this challenge. 

Motivated by the problem mentioned above, 
we propose further constraints on the states and inputs as follows:
\begin{equation}\label{eq:sch_error_z&u}
\begin{aligned}
-(e^{z_{\rm max}} + \epsilon^{z}_{i|k})  \leq (z_{i|k} - \hat{z}_{i|k}) \leq (e^{z_{\rm max}} + \epsilon^{z}_{i|k}), \\
-(e^{u_{\rm max}} + \epsilon^{u}_{i|k})\leq (u_{i|k} - \hat{u}_{i|k}) \leq (e^{u_{\rm max}} + \epsilon^{u}_{i|k}),
    \end{aligned}
\end{equation} 
$z_{i|k}$ and $u_{i|k}$ are the estimated states and inputs at step $i$, time $k$, while, $\hat{z}_{i|k}$ and $\hat{u}_{i|k}$ are predicted states and inputs that were calculated at step $i+1$, time $k-1$ as follows
\begin{equation*}
    \hat{z}_{i|k} = z^*_{i+1|k-1}, \quad \hat{u}_{i|k} = u^*_{i+1|k-1}.
\end{equation*}
Also,
$e^{z_{\rm max}}$ and $e^{u_{\rm max}}$ represent the bounds for the error between predicted and estimated states and inputs, i.e., $\hat{z}_{i|k},\hat{u}_{i|k}$ and $z_{i|k}, u_{i|k}$, respectively. Furthermore,  $\epsilon^{z}_{i|k}$ and $\epsilon^{u}_{i|k}$ are slack variables. 
Due to the introduction of the slack variables, the constraint~\eqref{eq:sch_error_z&u} can be seen as a soft constraint. Also, based on the states and inputs that the scheduling parameter depends on, we can only bound those states and inputs instead of all of the states and inputs.

The enforcement of~\cref{eq:sch_error_z&u} in the LPVMPC requires the optimization problem to find the estimated inputs and states, i.e., $z_{i|k}, u_{i|k}$, in a way that they remain close to the predicted values from the previous time instant, i.e., $\hat{z}_{i|k},\hat{u}_{i|k}$. Additionally, slack variables are used to allow the deviation of $z_{i|k}, u_{i|k}$ from $\hat{z}_{i|k},\hat{u}_{i|k}$ as long as the optimization problem remains feasible. 
Therefore, we propose the following concept.
\begin{definition}[\textbf{\textit{Scheduling trust region}}]\label{def:sch_TR}
    The region built by enforcing the constraints in~\eqref{eq:sch_error_z&u} in LPVMPC is called the scheduling trust region, i.e., when those constraints are satisfied, then $\hat{p}_{i|k} = p^*_{i+1|k-1}$ can be trusted to compute the system matrices at step $i$, time $k$, as $A(\hat{p}_{i|k})$ and $B(\hat{p}_{i|k})$. This constraint prevents the infeasibility of the LPVMPC, which can happen due to the modeling error that comes from poor scheduling parameter predictions.
\end{definition}

Since the constraint~\eqref{eq:sch_error_z&u} is a soft constraint, we do not require the exact values of $e^{z_{\rm max}}$ and $e^{u_{\rm max}}$.
In case the bound is too tight, the LPVMPC can use the slack variable to avoid infeasibility. One way to get a reasonable guess for $e^{z_{\rm max}}$ and $e^{u_{\rm max}}$, is to
run the nonlinear and LPV models and compare the LPV model's estimations to the actual nonlinear system. 
Also, we might have a guess for the values of $e^{z_{\rm max}}$ and $e^{u_{\rm max}}$ based on experience or prior knowledge about the rate of change of the scheduling parameter. Or, by tuning, we can try to find the values that lead to the best performance for the system. 
It is also important to note that if the scheduling trust region constraints in~\eqref{eq:sch_error_z&u} are only defined by the slack variables $\epsilon^{z}_{i|k}$ and $\epsilon^{u}_{i|k}$, then those constraints might attempt to keep the error $(z_{i|k} - \hat{z}_{i|k})$ and $(u_{i|k} - \hat{u}_{i|k})$ close to zero. This, in turn, can result in a significant loss of performance.

\section{Modeling}\label{Sec:Models}
In this paper, a dynamic bicycle model with linear tire forces represents the vehicle model that will be used in the NMPC. 

\subsection{Nonlinear Vehicle Model}\label{SubSec:NModel}
\begin{figure}
    \centering
    \includegraphics[scale=0.6]{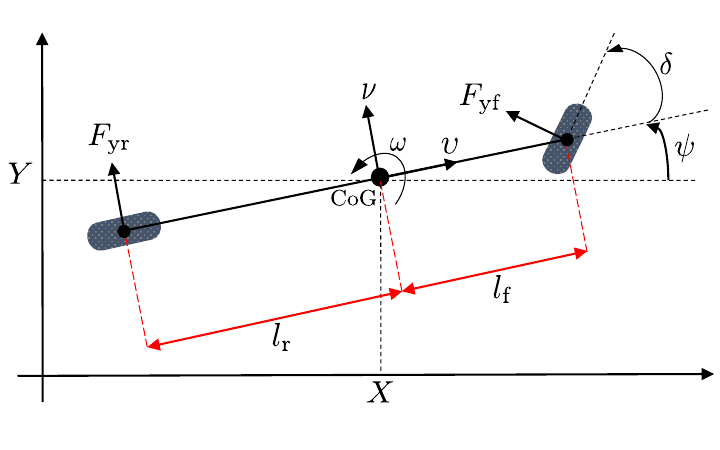}
    \caption{Vehicle dynamics representation~\cite{nezami2023design}}
    \label{fig:car_model}
\end{figure}

Based on~\cite[p.~27]{Rajamani2012}, the continuous differential equations describing the motion at time $t\geq 0$ of a vehicle are presented as follows
\begin{subequations}\label{eq:dynamic_model}
\begin{align}
    \dot{X}(t) &= \upsilon(t)\cos{\psi(t)}-\nu(t)\sin{\psi(t)},  \label{eq:Xdot}\\
    \dot{Y}(t) &= \upsilon(t)\sin{\psi(t)}+\nu(t)\cos{\psi(t)}, \label{eq:Ydot} \\
    \dot{\upsilon}(t) &= \omega(t) \nu(t) + a(t), \label{eq:xdotdot}\\
    \dot{\nu}(t) &= -\omega(t)\upsilon(t) + \frac{2}{m}( F_{\rm yf}(t) \cos{\delta(t)} + F_{\rm yr}(t)), \label{eq:ydotdot}\\
    \dot{\psi}(t) &= \omega(t), \label{eq:Psidot} \\
    \dot{\omega}(t) &= \frac{2}{I_{\rm z}} ( l_{\rm f} F_{\rm yf}(t) - l_{\rm r}  F_{\rm yr}(t)),\label{eq:psidotdot}
\end{align}
\end{subequations}
where $X$, $Y$, $\upsilon$, $\nu$, $\psi$ and $\omega$ denote the global $X$ axis coordinate of the center of gravity (GoG), the global $Y$ axis coordinate of the CoG, the longitudinal speed in body frame, the lateral speed in body frame, the vehicle yaw angle and the yaw angle rate, respectively. See~\cref{fig:car_model} for illustration. 
The control inputs of the system are the longitudinal acceleration $a$ and the steering angle $\delta$. The vehicle moment of inertia and mass are denoted by $I_{\rm z}$ and $m$, respectively. The lateral forces acting on the front and rear tires are denoted as $F_{\rm yf}$ and $F_{\rm yr}$, respectively, and calculated linearly as $F_{\rm yf} = C_{\alpha \rm f}  \alpha_{\rm f}$, $F_{\rm yr} = C_{\alpha \rm r}  \alpha_{\rm r}.$
The parameters $C_{\alpha \rm f}$ and $C_{\alpha \rm r}$ represent the cornering stiffness of the front and rear tires, respectively. The slip angle of the front tire is $\alpha_{\rm f}$ and is calculated as $\alpha_{\rm f} = \delta - (\nu + l_{\rm f} \omega)/\upsilon$. The rear tire slip angle is $\alpha_{\rm r}$ and is calculated as $\alpha_{\rm r} = (l_{\rm r} \omega -\nu )/\upsilon$. Moreover, the parameters used in this paper are given in~\cref{tab:VarPar}.

\begin{table}
    \centering
    \caption{Vehicle parameters~\cite{nezami2023design}}
    \label{tab:VarPar}
    \resizebox{\columnwidth}{!}{\begin{tabular}{lll}
        \toprule
        \rowcolor{mydarkgray} \textbf{Symbol}   & \textbf{Variables} & \textbf{Unit} \\ 
                           $X$ & Global X-axis coordinates of the vehicle's CoG & m \\
        \rowcolor{mygray} $Y$ & Global Y-axis coordinates of the vehicle's CoG & m  \\
        $\upsilon$ & Longitudinal velocity of the vehicle & $\rm m/s$ \\
        \rowcolor{mygray} $\nu$ & Lateral velocity of the vehicle & $\rm m/s$ \\
         $\psi$ & Yaw angle of the vehicle & $\rm rad$  \\
       \rowcolor{mygray}  $\omega$  & Yaw rate of the vehicle & $\rm rad/s$  \\
         $\delta$ & Steering angle of the front tire & $\rm rad$ \\
       \rowcolor{mygray}  $a$ & Longitudinal acceleration of the vehicle & $\rm m/s^2$ \\
         $\alpha_{\rm f}$ & Front tire slip angle  & $\rm rad$ \\
       \rowcolor{mygray} $\alpha_{\rm r}$ & Rear tire slip angle  & $\rm rad$ \\ \toprule
	  \rowcolor{mydarkgray}    \textbf{Symbol}   & \textbf{Parameter} & \textbf{Value/Unit}  \\
         $C_{\alpha \rm f}$ & Cornering stiffness front tire & $156~\rm kN/rad$  \\
        \rowcolor{mygray}  $C_{\alpha \rm r}$ & Cornering stiffness rear tire & $193~\rm kN/rad$ \\
        $l_{\rm f}$ & Distance CoG to front axle & $1.04~\rm m$ \\
        \rowcolor{mygray} $l_{\rm r}$ & Distance CoG to rear axle & $1.4~\rm m$  \\
        $I_{\rm z}$ & Vehicle yaw inertia & $2937~\rm kgm^2$ \\
        \rowcolor{mygray} $m$ & Vehicle mass & $1919~\rm kg$  \\
        $t_{\rm s}$  & sampling time & $0.05~\rm s$ \\
    \end{tabular}}
\end{table}

To utilize the model in~\cref{eq:dynamic_model} in an MPC framework, it is necessary to discretize it. One of the commonly used methods for obtaining the corresponding  discrete-time system is the forward Euler method, as follows
\begin{equation*}
    \dot{z}(t_k)\approx\frac{z(t_k+t_{\rm s})-z(t_k)}{t_{\rm s}}, \qquad t_k=t_{\rm s}k,~k=0,1,\ldots. 
\end{equation*}
In this way, 
the vehicle dynamics in~\cref{eq:dynamic_model} can be written as 
\begin{equation}\label{eq:dis_model}
z_{k+1} = z_k + t_{\rm s} f(z_k,u_k),
\end{equation}
where $z_k = \begin{bmatrix} X_k & Y_k &\upsilon_k & \nu_k & \psi_k & \omega_k \end{bmatrix}^\top$, $u_k = \begin{bmatrix} \delta_k & a_k \end{bmatrix}^\top$ with the sampling time, i.e., $t_{\rm s}$, given in Table~\ref{tab:VarPar}. The function, $f(z_k,u_k) = \dot{z}(z_k,u_k)$ is computed based on the differential equations in~\eqref{eq:dynamic_model}. 

\subsection{LPV Model}
By introducing the scheduling parameters $\upsilon(t)$, $\nu(t)$, $\delta(t)$, and $\psi(t)$, the scheduling parameter vector is defined as $p(t)= \begin{bmatrix}
    \upsilon(t) & \nu(t)& \delta(t) &\psi(t)
\end{bmatrix}^\top.$ 
The continuous-time nonlinear dynamics in~\cref{eq:dynamic_model} can be written equivalently in the LPV representation as
\begin{equation}\label{eq:dis_model}
\left\{\begin{aligned}
    \dot{z}(t)&=A_c(p(t))z(t)+B_c(p(t))u(t),\\
    p(t)&=(\upsilon(t),\nu(t),\delta(t),\psi(t)),~t\geq 0,
    \end{aligned}\right.
\end{equation}
where the subscript ``c" indicates the continuous in-time operator. The state vector $z(t)$ of dimension $6$ can be defined as $z(t):=\left[\begin{array}{cccccc}
    X(t) & Y(t) & \upsilon(t) & \nu(t) & \psi(t) & \omega(t)\end{array}\right]^\top$ with initial conditions $\bar{z}_0$ and the continuous-time system matrices $A_c\in\IR^{6\times 6},~B_c\in\IR^{6\times 2}$ as
\begin{equation*}
\begin{aligned}
    A_c(p(t))&:=\left[\begin{array}{cccccc}
       \! 0 \! & \! 0 \! & \! \cos(\psi(t)) \! & \! -\sin(\psi(t)) \! & \! 0 \! & \! 0 \! \\
       \! 0 \! & \! 0 \! & \! \sin(\psi(t)) \! & \! +\cos(\psi(t)) \! & \! 0 \! & \! 0 \! \\
       \! 0 \! & \! 0 \! & \! 0 \! & \! 0 \! & \! 0 \!  & \! \nu(t) \!\\
       \! 0 \! & \! 0 \! & \! 0 \! & \! a_{44}(t) \! & \! 0 \! & \! a_{46}(t) \! \\
       \! 0 \! & \! 0 \! & \! 0 \! & \! 0 \! & \! 0 \! & \! 1\! \\
       \! 0 \! & \! 0 \! & \! 0 \! & \! a_{64}(t) \! & \! 0 \! & \! a_{66}(t) \!\end{array}\right],
        \end{aligned}
\end{equation*}
\begin{equation*}
    \begin{aligned}
         \beta_f:=&\frac{2C{af}}{m},\beta_r:=\frac{2C{ar}}{m},~\gamma_f:=\frac{2\ell_fC_{af}}{I_z},\gamma_r:=\frac{2\ell_rC_{ar}}{I_z},
    \end{aligned}
\end{equation*}
\begin{equation*}
    \begin{aligned}
        a_{44}(t)&:=-\beta_f\cos(\delta(t))\frac{1}{\upsilon(t)}-\beta_r\frac{1}{\upsilon(t)},\\
        a_{46}(t)&:=-\upsilon(t)-\beta_f\cos(\delta(t))\frac{1}{\upsilon(t)}\ell_f+\beta_r\frac{1}{\upsilon(t)}\ell_r,\\
        a_{64}(t)&:=\frac{1}{\upsilon(t)}(\gamma_r-\gamma_f),~a_{66}(t):=-\frac{1}{\upsilon(t)}(\gamma_f\ell_f+\gamma_r\ell_r),
    \end{aligned}
\end{equation*}
and
\begin{equation*}
\small
    \begin{aligned}
       B_c(p(t)):= \begin{bmatrix}
           0 & 0 & 0 &  \beta_f\cos(\delta(t)) & 0 & \gamma_f \\
       0 & 0 &  1 &0 & 0 & 0 
       \end{bmatrix} ^\top.
    \end{aligned}
\end{equation*}
The discretization with the Euler method and the sampling time $t_{\rm s}$ results in the discrete-time LPV representation of  \cref{eq:dis_model} as
\begin{equation}\label{eq:dis_LPV}
     \left\{\begin{aligned}      z_{k+1}&=A(p_k)z_k+B(p_k)u_k,\\
      p_k&=(\upsilon_k,\nu_k,\delta_k,\psi_k),~k\in\mathbb{Z}_+\cup \{ 0\}
      \end{aligned}\right.
\end{equation}
where $A(p_k)=\mathbb{I}_6+t_{\rm s} A_c(p_k),~B(p_k)=t_{\rm s} B_c(p_k)$ are the discrete-time LPV system matrices.

\section{Obstacle avoidance with NMPC}\label{sec:NMPC}
This section focuses on introducing an NMPC framework for obstacle avoidance while following a reference trajectory with respect to vehicle constraints and road boundaries. In this reference tracking problem, whenever a new obstacle appears on the road, it is the controller's responsibility to determine the most appropriate time and maneuver to overtake the obstacle and then return the vehicle to the reference.

We also assume that the other traffic participants are ellipses. This is a common assumption, as the surrounding obstacles are usually other vehicles, and the area occupied by vehicles can be presented with an ellipse conveniently~\cite{muraleedharan2021real}.

The constrained nonlinear optimal control problem for obstacle avoidance while following a reference trajectory is formulated as follows
\begin{subequations}\label{eq:nonlinear_MPC}
	\begin{align} 
		\underset{U}{\text{min}} \
		& \| z_{N|k} \! -\! z^{\text{ref}}_{N|k} \|^2_P\! + \!\sum_{i=0}^{N-1}\left(\| z_{i|k} - z^{\text{ref}}_{i|k} \|^2_Q +  \| u_{i|k}  \|^2_R\right)   \\
		 \text{s.t.} \;\;
		& z_{i+1|k} \!= \!z_{i|k} \!+\! t_s f(z_{i|k},u_{i|k}),  \forall i \!= \!0,\ldots,N\!-\!1,  \label{eq:NMPC_model} \\
  		& z_{0|k} = z_k, \label{eq:NMPC_initial_condition} \\
            &  u_{i|k} \in \mathcal{U}, \qquad \forall i = 0,1,\ldots,N-1, \label{eq:NMPC_input_cons} \\
		&  z_{i|k} \in \mathcal{Z}, \quad \forall i = 0,1,\ldots,N,  \label{eq:NMPC_state_cons}\\
            & \frac{( X_{\text{obs}} - X_{i|k})^2}{r_x^2} + \frac{(Y_{\text{obs}} - Y_{i|k})^2}{r_y^2} \geq 1, \label{eq:NMPC_obs_cons} \\
            &  \begin{bmatrix}
        a_{1,i|k} & b_{1,i|k} \\ 
        -a_{2,i|k} & -b_{2,i|k}
    \end{bmatrix} \begin{bmatrix}
        X_{i|k} \\ Y_{i|k}
    \end{bmatrix} \leq \begin{bmatrix}
        c_{1,i|k} \\ -c_{2,i|k}
    \end{bmatrix}, \label{eq:NMPC_road_cons}
	\end{align}
\end{subequations}
where, the tuning matrices are $Q \succeq 0 \in \mathbb{R}^{6 \times 6}$, $R \succ 0 \in \mathbb{R}^{2 \times 2}$ and $P \succeq 0 \in \mathbb{R}^{6 \times 6}$. 
The input constraint $\mathcal{U}$ in~\eqref{eq:NMPC_input_cons} and the state constraint $\mathcal{Z}$ in~\eqref{eq:NMPC_state_cons} are defined in~\cref{eq:prel_input_state_cons}, where $G^u = \begin{bmatrix}
    \mathbb{I}_2 & -\mathbb{I}_2
\end{bmatrix}^\top$, $h^u = \begin{bmatrix}
    \delta_{\rm max} & a^{\rm lon}_{\rm max} & \delta_{\rm max} & a^{\rm lon}_{\rm max}
\end{bmatrix}^\top$, $G^z = \begin{bmatrix}
    \mathbb{I}_6 & -\mathbb{I}_6
\end{bmatrix}^\top$ and $h^z = \begin{bmatrix}
    h^z_{\rm max} & h^z_{\rm max}
\end{bmatrix}^\top$, with $h^z_{\rm max} = \begin{bmatrix}
    X_{\rm max} & Y_{\rm max} & \upsilon_{\rm max} & \nu_{\rm max} & \psi_{\rm max} & \omega_{\rm max}
\end{bmatrix}^\top$. In~\eqref{eq:NMPC_obs_cons}, $(X_{\text{obs}},Y_{\text{obs}})$ represent the center of an ellipsoidal obstacle with axes $r_x, r_y$. The enforcement of constraint~\eqref{eq:NMPC_obs_cons} in the NMPC guarantees the vehicle stays outside of the region occupied by the obstacle. The constraint~\eqref{eq:NMPC_road_cons} is the road boundary constraint, which the value of  $a_{1,i|k}$, $b_{1,i|k}$, $a_{2,i|k}$, $b_{2,i|k}$, $c_{1,i|k}$ and $c_{2,i|k}$ are computed based on the tangent line to the road boundary at each side of the road at time step $i+k$.

The vector $z^{\text{ref}}_{i|k} = \begin{bmatrix} X_{i|k}^{\rm ref} & Y_{i|k}^{\rm ref} & \upsilon_{i|k}^{\rm ref} & \nu_{i|k}^{\rm ref} & \psi_{i|k}^{\rm ref} & \omega_{i|k}^{\rm ref} \end{bmatrix}^\top$ is the reference value for the states at time step $i+k$, which is computed as explained below. 

We assume that only the $(X_k^{\rm ref}, Y_k^{\rm ref})$ values of the reference trajectory are available, where $X_k^{\rm ref}$ is the global X axis coordinate and $Y_k^{\rm ref}$ is the global Y axis coordinate of a reference point. Then, we compute the corresponding reference values for the remaining four states, $\upsilon_k^{\rm ref}$, $\nu_k^{\rm ref}$, $\psi_k^{\rm ref}$ and $\omega_k^{\rm ref}$, to track the reference trajectory effectively. 
For the computation of $\psi_k^{\rm ref}$, the global $(X_k^{\rm ref}, Y_k^{\rm ref})$ can be directly used as follows
\begin{equation}\label{eq:psi_ref}
    \psi_k^{\rm ref} = \arctan\left(\frac{Y_{k-1}^{\rm ref} - Y_{k}^{\rm ref}}{X_{k-1}^{\rm ref} - X_{k}^{\rm ref}}\right). 
\end{equation}
Next, $\omega_k^{\rm ref}$ can be calculated as $\omega_k^{\rm ref} = (\psi_k^{\rm ref} - \psi_{k-1}^{\rm ref})/t_s$, where $\psi_{k}^{\rm ref}$ and $\psi_{k-1}^{\rm ref}$ are the reference yaw angles which are computed in the previous step by using~\cref{eq:psi_ref}. To calculate $\upsilon_k^{\rm ref}$ and $\nu_k^{\rm ref}$, which represent the longitudinal and lateral speeds in the body frame, the reference points in the body frame are 
determined as follows 
\begin{equation}\label{eq:glob_to_loc}
    \begin{bmatrix}
        x_k^{\rm ref}  \\
        y_k^{\rm ref}
    \end{bmatrix} \!=\! \begin{bmatrix}
        \cos{(\psi_k^{\rm ref} )} & \sin{(\psi_k^{\rm ref}) } \\
        - \sin{(\psi_k^{\rm ref} )} & \cos{(\psi_k^{\rm ref} )}
    \end{bmatrix} \begin{bmatrix}
        X_{k}^{\rm ref} - X_{k-1}^{\rm ref}  \\ Y_{k}^{\rm ref} - Y_{k-1}^{\rm ref}
    \end{bmatrix}. 
\end{equation}
Then, 
the reference speeds
can be readily computed as $\upsilon_k^{\rm ref} = (x_k^{\rm ref} - x_{k-1}^{\rm ref})/t_s$ and $\nu_k^{\rm ref} = (y_k^{\rm ref}-y_{k-1}^{\rm ref})/t_s$, where $x_i^{\rm ref}$ and $y_i^{\rm ref}$, for $i=k,k-1$, are computed in~\cref{eq:glob_to_loc}.


\section{Obstacle avoidance with LPVMPC}\label{sec:LPVMPC}
This section aims to propose a novel LPVMPC to achieve the same goal as the NMPC in the previous section. Prior to introducing LPVMPC, we present the linearization of the nonlinear constraints that will be utilized within the LPVMPC framework.  

\subsection{Linearization of Obstacle Avoidance Constraint} \label{Subsec:linobs}

To make sure the vehicle avoids the obstacle, the constraint~\eqref{eq:NMPC_obs_cons} should be satisfied. 
However, the constraint~\eqref{eq:NMPC_obs_cons} is a nonlinear constraint that cannot be directly used in the quadratic programming of an LPVMPC.
To address this problem, this paper proposes a method for linearizing this constraint, as presented below. 

Given that the reference trajectory is presented as a series of points, whenever a reference point falls within the obstacle for a step over the MPC horizon, we propose linearization of the nonlinear constraint~\eqref{eq:NMPC_obs_cons} around that particular reference point. Subsequently, the nonlinear constraint~\eqref{eq:NMPC_obs_cons} can be replaced with only a halfspace constraint, i.e., the vehicle is forced to stay within a halfspace that does not contain the obstacle on that time step. The linear constraint has to change adaptively when a new reference point is found inside the obstacle. 

 \begin{assumption}\label{assum_1}
In this approach, it is important to emphasize that there is an assumption regarding having a predetermined decision of overtaking an obstacle from a specific side. This can be done by using any planners that are available for obstacle avoidance, e.g., topology-based planners focus on determining the side on which a robot should overtake an obstacle~\cite{de2023globally}.
 \end{assumption}

By defining $a_{3,k}, b_{3,k}$ and $c_{3,k}$ as the parameters of the tangent half-space to the obstacle, the $(X_k,Y_k)$ coordinate of the vehicle should then satisfy the following linear inequality 
\begin{equation} \label{e:lin_cons}
    h_1 : a_{3,k} X_k + b_{3,k} Y_k \geq c_{3,k}. 
\end{equation}
The above inequality is the obstacle avoidance constraint which is enforced in the LPVMPC~\eqref{eq:LPV_MPC} as a state constraint. 

\begin{figure}
    \centering
    \includegraphics[scale=0.8]{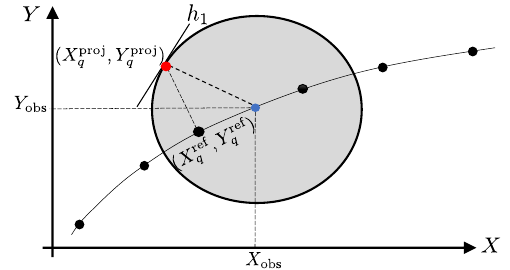}
    \caption{\centering Linear obstacle avoidance constraint}
    \label{fig:lin_obs}
\end{figure}

Assume that a point on the reference trajectory, e.g., $(X_q^{\rm ref}, Y_q^{\rm ref})$ falls inside the obstacle over the MPC horizon. Find the point $Q=(X_q^{\rm proj},Y_q^{\rm proj})$ as the projection of $(X_q^{\rm ref}, Y_q^{\rm ref})$ on the circumference of the obstacle with respect to the curvature of the road, as depicted in~\cref{fig:lin_obs}. 
Define: $G(X_k,Y_k)=r_y^2(X_k-X_{\text{obs}})^2+r_x^2(Y_k-Y_{\text{obs}})^2-r_x^2r_y^2$. 
The gradient is:
\begin{equation}
    \nabla G(X_k,Y_k)=\left[\begin{array}{cc}
       2r_y^2(X_k-X_{\text{obs}})  &  2r_x^2(Y_k-Y_{\text{obs}})\\
    \end{array}\right].
\end{equation}
Thus, the tangent line at point $Q=(X_q^{\rm proj},Y_q^{\rm proj})$ is
\begin{equation*}
\begin{aligned}
    0&=\underbrace{G(X_q^{\rm proj},Y_q^{\rm proj})}_{0}+\nabla G(X_q^{\rm proj},Y_q^{\rm proj})\left[\begin{array}{c}
         X_k-X_q^{\rm proj}  \\
         Y_k-Y_q^{\rm proj} 
    \end{array}\right],
    \end{aligned}
\end{equation*}
which is equivalent to the following equation
\begin{equation*}
\begin{aligned}
     0 = &2r_y^2(X_q^{\rm proj}-X_{\text{obs}})(X_k-X_q^{\rm proj})+ \\
       &2r_x^2(Y_q^{\rm proj}-Y_{\text{obs}})(Y_k-Y_q^{\rm proj}),
\end{aligned}
\end{equation*}
which leads to 
\begin{equation}\label{eq:tangent_param}
\begin{aligned}
     &\underbrace{r_y^2(X_q^{\rm proj}-X_{\text{obs}})}_{a_{3,k}}X_k+\underbrace{r_x^2(Y_q^{\rm proj}-Y_{\text{obs}})}_{b_{3,k}}Y_k=\\
  &=\underbrace{r_y^2(X_q^{\rm proj}-X_{\text{obs}})X_q^{\rm proj}+r_x^2(Y_q^{\rm proj}-Y_{\text{obs}})Y_q^{\rm proj}}_{c_{3,k}},
\end{aligned}
\end{equation}
where, $a_{3,k}, b_{3,k}$ and $c_{3,k}$ are the parameters of the tangent half space in~\cref{e:lin_cons}. When there is no obstacle in the MPC horizon, $c_{3,k}$ is equal to $-\infty$.
The algorithm for the suggested method is presented in~\cref{alg:lin_obs}. 

\begin{algorithm}
\caption{Adaptive linear obstacle avoidance constraint}\label{alg:lin_obs}
\begin{algorithmic}
\State $k = 0, 1,2, \cdots$;
\State Points $(X_q^{\rm ref}, Y_q^{\rm ref})$ are inside the obstacle
\If{$(X_q^{\rm ref}, Y_q^{\rm ref})$ appears in the MPC horizon}
    \State Find the projection of the point $(X_q^{\rm ref}, Y_q^{\rm ref})$ on the radius of the obstacle, i.e., $(X_q^{\rm proj},Y_q^{\rm proj})$; \\
    \State Find parameters of the tangent line as presented in~\cref{eq:tangent_param}; \\
    \State Enforce the constraint  $h_1 : a_{3,k} X_k + b_{3,k} Y_k \geq c_{3,k}$;
\EndIf
\end{algorithmic}
\end{algorithm}

\subsection{LPVMPC}
The suggested LPVMPC is as follows
\begin{subequations}\label{eq:LPV_MPC}
	\begin{align} 
		\underset{U, E}{\text{min}} \
		& \! \| z_{N|k} \!\! -\! \!z^{\text{ref}}_{N|k} \|^2_P\! + \!\!\! \sum_{i=0}^{N-1}\! \! \! \left( \!  \| z_{i|k} \!- \!z^{\text{ref}}_{i|k} \|^2_Q \! + \! \|\! u_{i|k}  \!\|^2_R  \! + \! \| \! \epsilon_{i|k}  \! \|^2_{E_p} \!\! \right)~\label{eq:LPV_cost}  \\
		 \text{s.t.} \;\;
		& \!z_{i+1|k}\!=\!A(\hat{p}_{i|k})z_{i|k}\!\!+\!\!B(\hat{p}_{i|k})u_{i|k},~i\!\!=\!\!0,\!\ldots\!,\!N\!\!-\!\!1,\!\!\!\!\!\!\label{eq:LPVMPC_model} \\
		& z_{0|k} = z_k, \label{eq:LPVMPC_initial_condition} \\
            & u_{i|k} \in \mathcal{U}, \qquad \forall i = 0,1,\ldots,N-1, \label{eq:LPVMPC_input_cons} \\
            &  z_{i|k} \in \mathcal{Z}, \quad \forall i = 0,1,\ldots,N,  \label{eq:LPVMPC_state_cons}\\
                &a_{3,i|k} X_{i|k} + b_{3,i|k} Y_{i|k} \geq c_{3,i|k}, \label{eq:LPVMPC_obs_cons}\\
             &  \begin{bmatrix}
        a_{1,i|k} & b_{1,i|k} \\ 
        -a_{2,i|k} & -b_{2,i|k}
         \end{bmatrix} \begin{bmatrix}
        X_{i|k} \\ Y_{i|k}
        \end{bmatrix} \leq \begin{bmatrix}
        c_{1,i|k} \\ -c_{2,i|k}
     \end{bmatrix}, \label{eq:LPVMPC_road_cons} \\
   & -(e^{z_{\rm max}} \! + \! \epsilon^{z}_{i|k})  \! \leq \! (z_{i|k} \! - \! \hat{z}_{i|k})\!  \leq \! (e^{z_{\rm max}}\!  + \!\epsilon^{z}_{i|k}), \label{eq:LPVMPC_smooth_x} \\
& -(e^{u_{\rm max}}\! + \!\epsilon^{u}_{i|k})\!\leq\! (u_{i|k} - \hat{u}_{i|k})\! \leq\! (e^{u_{\rm max}} \!+ \!\epsilon^{u}_{i|k}), \label{eq:LPVMPC_smooth_u}
	\end{align}
\end{subequations}
where the reference trajectory $z^{\text{ref}}_{i|k}$, the tuning matrices $P$, $Q$, $R$ are as defined for the Problem~\eqref{eq:nonlinear_MPC}.The decision variables are $U$ as for~\eqref{eq:nonlinear_MPC} and $E = \begin{bmatrix}
    \epsilon_{0|k} & \ldots & \epsilon_{N-1|k}
\end{bmatrix}$, with  $\epsilon_{i|k} = \begin{bmatrix}
    \epsilon^{z}_{i|k} & \epsilon^{u}_{i|k}
\end{bmatrix}^\top$. Also, the tuning matrix $E_p$ is $E_p \succeq 0 \in \mathbb{R}^{4 \times 4}$.
The LPV model in~\cref{eq:LPVMPC_model} is defined in~\cref{eq:dis_LPV}. 
The constraints~\eqref{eq:LPVMPC_input_cons} and~\eqref{eq:LPVMPC_state_cons} are as defined in~\eqref{eq:NMPC_input_cons} and~\eqref{eq:NMPC_state_cons}, respectively. The constraint~\eqref{eq:LPVMPC_obs_cons} is the linearized obstacle avoidance constraint, as explained in~\cref{Subsec:linobs}, which is active only when the obstacle is in the MPC horizon. 
The constraint~\eqref{eq:LPVMPC_road_cons} is added to make sure the vehicle moves between road boundaries, as in~\eqref{eq:NMPC_road_cons}. Finally, the soft constraints~\eqref{eq:LPVMPC_smooth_x} and~\eqref{eq:LPVMPC_smooth_u} are the scheduling trust region constraints due to the reasons explained in~\cref{Subsec:smoothschedul}.


\begin{algorithm}\label{algo:lpvmpc}
\caption{The QP-based LPVMPC algorithm}
\textbf{Input}: Initial conditions $z_k$, and the road reference $(X_k,Y_k),~k\in\mathbb{Z}_+$.\\
\textbf{Output}: The control input $u_k,~k=1,\ldots$, that drives the nonlinear system to the reference while avoiding obstacles.
\begin{algorithmic}[1]\label{alg:LPVMPC}
\State Initialize for $k=0$ the scheduling parameter vector $\hat{p}_{i|0}$ as
$$\hat{p}_{i|0}:=\left(\upsilon_0,~\nu_0,~\delta_0,~\psi_0\right),~i=0,\ldots,N-1$$ 
\While{$k=0,1,\ldots$} 
\State Update the state $z_{i|k}^{\text{ref}}$ as explained in~\cref{sec:NMPC}.
\State Solve the QP in Problem~\ref{eq:LPV_MPC}
\begin{equation*}
\begin{aligned}    
\left[z_{i+1|k},u_{i|k}\right]&\leftarrow\texttt{QP}(\hat{p}_{i|k},z_k,z_{i|k}^{\text{ref}}),~i=0,\ldots,N-1\\
\text{Update}~\hat{p}_{i|k}&:=\left(\hat{\upsilon}_{i|k},~\hat{\nu}_{i|k},~u_{i|k},~\hat{\psi}_{i|k}\right),~i=0,\ldots,N\\
\end{aligned}
\end{equation*}
  \State Apply $u_k=u_{0|k}$ to the system
  \State Measure $z_{k+1}$
  \State Update $\hat{p}_{i|k+1}=\hat{p}_{i+1|k},~i=0,\ldots,N-1$
  \State $k\leftarrow k+1$
  \EndWhile
\end{algorithmic}
\end{algorithm}

\section{Results}\label{sec:results}
This section compares the application of the LPVMPC with trust region to standard LPVMPC and NMPC. To provide a fairly realistic representation of a vehicle, a nonlinear dual-track vehicle body with 3 degrees of freedom from Vehicle Dynamics Blockset in Matlab~\cite{MATLAB:2021} is used to represent the vehicle in simulations. The simulations are performed on a Dell Latitude 5590 laptop with an Intel(R) Core(TM) i7-8650U CPU and 16 GB of RAM. The simulations are conducted in Simulink~\cite{MATLAB}, where the MPCs are implemented using YALMIP~\cite{Lofberg2004}.
In the simulations, \textit{Ipopt} serves as the nonlinear solver~\cite{wachter2006implementation}, while the quadratic solver employed is the Matlab solver, \textit{quadprog}. Also, the optimality tolerance of \textit{Ipopt} is $10^{-4}$ and for the \textit{quadprog} is $10^{-6}$.

The list of constraints used in the LPVMPC and the NMPC are presented in~\cref{tab:MPCPar}. Also, to keep the movement of the vehicle smoother constrains the rate of change of $\delta_k$ and $a_k$, i.e., $|\delta_{k} - \delta_{k-1} | \leq 25\pi / 180 \text{ rad}$, $| a_{k} - a_{k-1} | \leq 1.5$~$\rm m / \rm s^2$ are enforced in the optimization problems as well. 

\begin{table}[h]
    \centering
    \caption{MPC Parameters}
    \label{tab:MPCPar}
    \setlength{\tabcolsep}{2pt}
    \begin{tabular}{l l | l l} 
        \toprule
        \rowcolor{mydarkgray} \textbf{Parameter}   & \textbf{Value}  & \textbf{Parameter}    & \textbf{Value}\\ 
                       Lower bound on  $X_k$ &   -1 \text{m} & Upper bound on  $X_k$ &   1500 \text{m}  \\
        \rowcolor{mygray} Lower bound on  $Y_k$ &   -600 \text{m} & Upper bound on  $Y_k$ &   800 \text{m} \\
        Lower bound on  $\upsilon_k$ &   1 $\frac{\rm m}{\rm s}$ & Upper bound on  $\upsilon_k$ &   100 $\frac{\rm m}{\rm s}$  \\
        \rowcolor{mygray} Upper bound on  $|\nu_k|$ &   10 $\frac{\rm m}{\rm s}$ & Upper bound on  $|\psi_k|$ &   $\pi \text{ rad}$ \\
            Upper bound on  $|\omega_k|$ &   $\frac{\pi}{3t_s}\frac{\rm rad}{\rm s}$ & Sampling time $t_s$ & $0.05 \text{ s}$   \\
        \rowcolor{mygray}  Upper bound on  $| \delta_k |$ &   $\frac{34\pi}{180} \text{ rad} $ & Sampling frequency $f_s$ & $20 \text{ Hz}$\\
         Lower bound on  $a_k$ &   $-6$ $ \frac{ \rm m}{ \rm s^2}$ & Upper bound on  $a_k$ &   $2$ $ \frac{ \rm m}{ \rm s^2}$
    \end{tabular}
\end{table}
\subsection{Comparison of LPVMPC with trust region to standard LPVMPC}
In order to compare the performance of a standard LPVMPC to the LPVMPC with trust region as proposed in~\eqref{eq:LPV_MPC}, we set 10 driving scenarios. 
The driving scenarios involve driving the vehicle on a circular road with an obstacle blocking part of the road.
In these scenarios, the obstacles are considered as circles for simplicity. They are positioned along various parts of the road and have radii ranging from 0.7 m to 1.4 m. The horizon of the LPVMPCs has been set to either 8 or 15.

\begin{figure}
     \centering
     \includegraphics[scale=0.6]{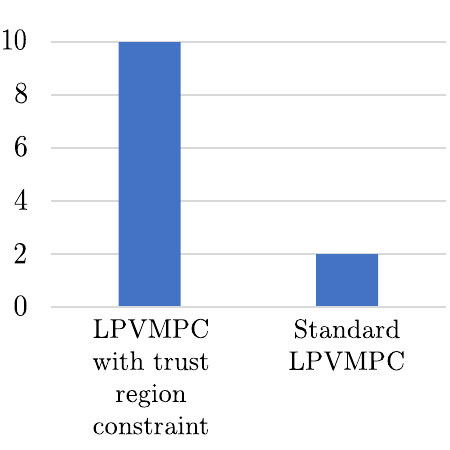}
        \caption{Comparison of the feasibility of the proposed LPVMPC with the trusts region in this paper to a standard LPVMPC for 10 obstacle avoidance scenarios.}
        \label{fig:LPV_comparison}
\end{figure}

\begin{figure}
     \centering
     \includegraphics[scale=0.7]{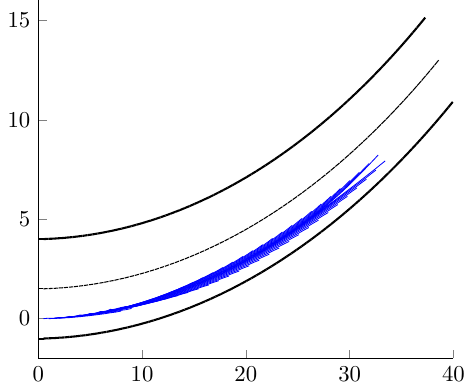}
        \caption{Open-loop trajectories of Standard LPVMPC at consecutive simulation time steps}
        \label{fig:openloop_standard_LPVMPC}
\end{figure}

\begin{figure}
     \centering
    \includegraphics[scale=0.7]{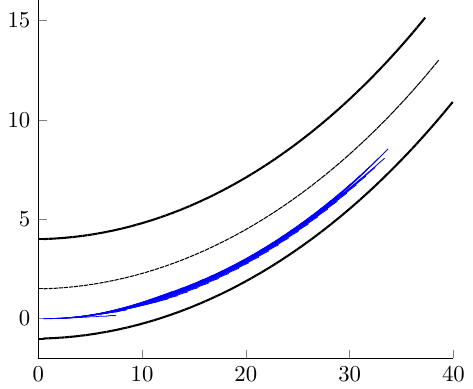}
        \caption{Open-loop trajectories of LPVMPC with trust region at consecutive simulation time steps}
        \label{fig:openloop_LPVMPC_trust_region}
\end{figure}

In these driving scenarios, it is observed that the standard LPVMPC is feasible only in 2 out of 10 driving scenarios. In the next step, we add the trust region constraint to the LPVMPC and observe that the optimization problem becomes feasible. This comparison is presented in~\cref{fig:LPV_comparison}. 
In~\cref{fig:openloop_standard_LPVMPC,fig:openloop_LPVMPC_trust_region}, the open-loop trajectories of one standard LPVMPC and the corresponding LPVMPC with trust region are presented.~\cref{fig:openloop_standard_LPVMPC} presents the open-loop trajectories of a standard LPVMPC, which shows a significant deviation in trajectories in two steps.
Since the scheduling parameter is a function of the states and inputs, this significant deviation leads to poor estimations by the LPV model. On the other hand,~\cref{fig:openloop_LPVMPC_trust_region} shows that adding the trust region constraint to the LPVMPC can fix this problem by keeping the open-loop trajectories closer to each other and consequently keeping the scheduling parameter prediction error smaller. However, the deviation of estimated states and inputs from the predicted states and inputs depends on the choice of $e^{z_{\rm max}}$ and~$ e^{u_{\rm max}}$ in~\cref{eq:LPVMPC_smooth_x} and~\cref{eq:LPVMPC_smooth_u}, as well as, the tuning matrix $E_p$ in the cost function~\eqref{eq:LPV_cost}. Therefore, the user can tune these parameters to get the desired performance for the vehicle while avoiding the infeasibility that comes from the scheduling parameter prediction error like in standard LPVMPC.  



\subsection{Comparison of LPVMPC with trust region to NMPC}

\subsubsection{Reference Tracking}
\begin{figure*}
     \centering
     \begin{subfigure}[b]{0.3\textwidth}
         \centering
          \includegraphics[scale=0.65]{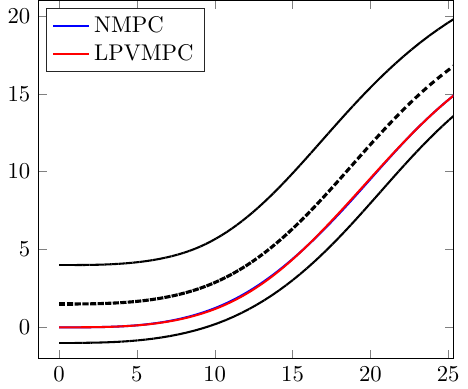}
  \caption{\centering XY position of the vehicle  }
    \label{fig:SC1_states}
     \end{subfigure}
     \hfill
     \begin{subfigure}[b]{0.3\textwidth}
         \centering
       \includegraphics[scale=0.65]{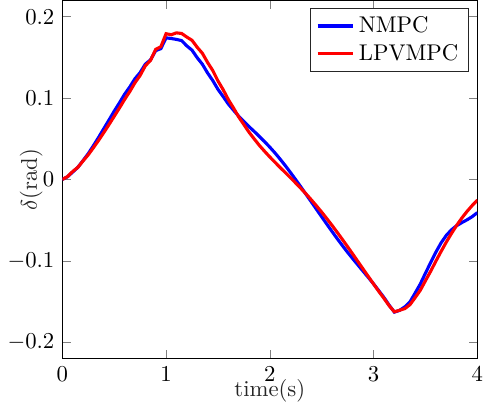}
  \caption{\centering Steering angle}
    \label{fig:SC1_delta}
     \end{subfigure}
     \hfill
     \begin{subfigure}[b]{0.3\textwidth}
         \centering
  \includegraphics[scale=0.65]{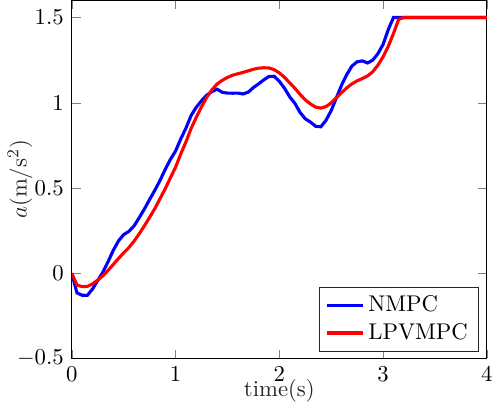}
  \caption{\centering Acceleration}
    \label{fig:SC1_a}
     \end{subfigure}
        \caption{First reference tracking scenario}
        \label{fig:SC1_refTrack}
\end{figure*}

At first, the performance of the NMPC and the LPVMPC are compared in two reference tracking scenarios. The result of the first scenario is depicted in~\cref{fig:SC1_refTrack}. In this scenario, the parameters $Q$ and $R$ for the NMPC are $Q = diag(\begin{bmatrix}
    10 & 10 & 5 & 1 & 1 & 1
\end{bmatrix})$ and  $R = diag(\begin{bmatrix}
    0.1 & 0.1
\end{bmatrix})$, while that for the LPV are $Q = diag(\begin{bmatrix}
    10 & 10 & 1 & 1 & 10 & 1
\end{bmatrix})$ and  $R = diag(\begin{bmatrix}
    0.1 & 0.1
\end{bmatrix})$. The horizon of both LPVMPC and NMPC is $8$. The values of $R_{1} = -1 $~m and $ R_{2} = 4 $~m represent the width of the road on each side of the vehicle.

\begin{figure*}
     \centering
     \begin{subfigure}[b]{0.3\textwidth}
         \centering
          \includegraphics[scale=0.65]{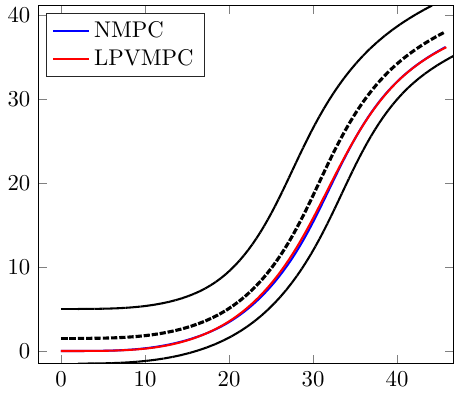}
  \caption{\centering XY position of the vehicle}
    \label{fig:SC2_states}
     \end{subfigure}
     \hfill
     \begin{subfigure}[b]{0.3\textwidth}
         \centering
         \includegraphics[scale=0.65]{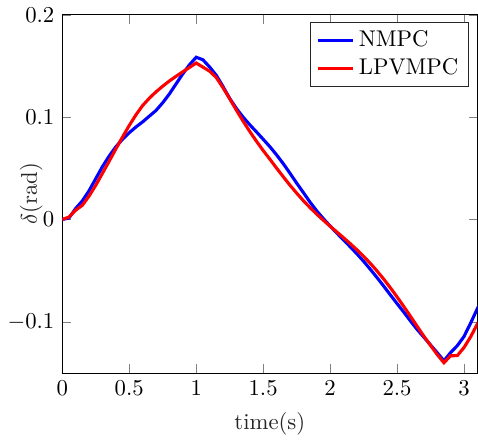}
  \caption{\centering Steering angle}
    \label{fig:SC2_delta}
     \end{subfigure}
     \hfill
     \begin{subfigure}[b]{0.3\textwidth}
         \centering
   \includegraphics[scale=0.65]{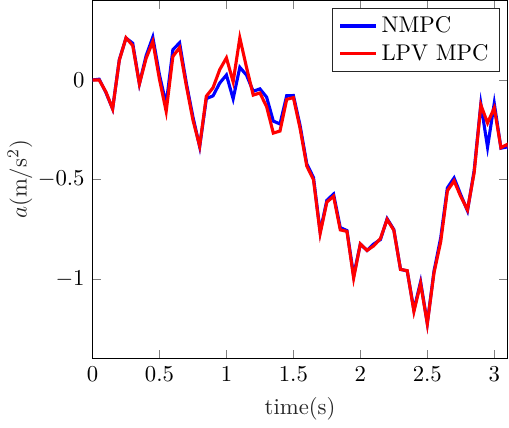}
  \caption{\centering Acceleration}
    \label{fig:SC2_a}
     \end{subfigure}
        \caption{Second reference tracking scenario}
        \label{fig:SC2_refTrack}
\end{figure*}

In the second reference tracking scenario, the parameters of NMPC are $Q = diag(\begin{bmatrix}
    10 & 10 & 1000 & 1 & 1 & 1
\end{bmatrix})$ and  $R = diag(\begin{bmatrix}
    0.001 & 0.001
\end{bmatrix})$. For the LPVMPC the parameters are $Q = diag(\begin{bmatrix}
    10 & 10 & 300 & 1 & 1 & 1
\end{bmatrix})$ and  $R = diag(\begin{bmatrix}
    0.001 & 0.001
\end{bmatrix})$. The horizon of both LPVMPC and NMPC is $8$. The value of road width in this scenario are $R_{1} = -1.5 $~m and $ R_{2} = 5 $~m. The result of this scenario comparison is presented in~\cref{fig:SC2_refTrack}.

\begin{table}[h!]
    \centering
    \caption{Comparison of computation times between NMPC and LPVMPC for the reference tracking scenario 1}
    \label{tab:SC1}
    \setlength{\tabcolsep}{2pt}
    \begin{tabular}{ l | l | l } 
        \toprule
        \rowcolor{mydarkgray} \textbf{RT w/o obstacle}   & \textbf{NMPC}  & \textbf{LPVMPC}  \\ \hline
                              Average time & 0.3067  s &  0.0145 s \\
        \rowcolor{mygray}    Maximum time &  1.2334 s  & 0.0225 s \\
                             Minimum time  & 0.0973 s  & 0.0121 s   
    \end{tabular}
\end{table}

\begin{table}[h!]
    \centering
    \caption{Comparison of computation times between NMPC and LPVMPC for the reference tracking scenario 2}
    \label{tab:SC2}
    \setlength{\tabcolsep}{2pt}
    \begin{tabular}{ l | l | l} 
        \toprule
        \rowcolor{mydarkgray} \textbf{RT w/o obstacle}   & \textbf{NMPC}  & \textbf{LPVMPC}  \\ \hline
                              Average time &  0.1982 s &  0.0155 s \\
        \rowcolor{mygray}    Maximum time &  0.3882 s  &  0.0335 s \\
                             Minimum time  & 0.1074 s  &  0.0132 s   
    \end{tabular}
\end{table}

The performance analysis depicted in~\cref{fig:SC1_refTrack} and~\cref{fig:SC2_refTrack} indicates that LPVMPC and NMPC have nearly identical performance regarding reference tracking error. However, comparing the inputs shows that NMPC generates smoother steering angles and accelerations. The computation times for solving the NMPC and the LPVMPC are detailed in~\cref{tab:SC1} and~\cref{tab:SC2}. The numbers in these tables indicate that the LPVMPC performs significantly faster than NMPC. This result aligns with expectations since NMPC needs to solve a nonlinear optimization problem, whereas LPVMPC is based on a quadratic optimization problem.

\subsubsection{Obstacle Avoidance}
The second part tests the LPVMPC and NMPC for reference tracking while overtaking an obstacle. The values of the parameters used in the LPVMPC and the NMPC in scenarios 1 and 2 are the same as the ones presented in the reference tracking section.

\begin{figure*}
     \centering
     \begin{subfigure}[b]{0.3\textwidth}
         \centering
          \includegraphics[scale=0.65]{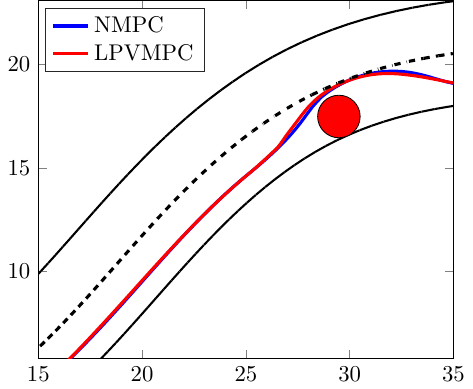}
  \caption{\centering XY position of the vehicle}
    \label{fig:SC1_obs_states}
     \end{subfigure}
     \hfill
     \begin{subfigure}[b]{0.3\textwidth}
         \centering
          \includegraphics[scale=0.65]{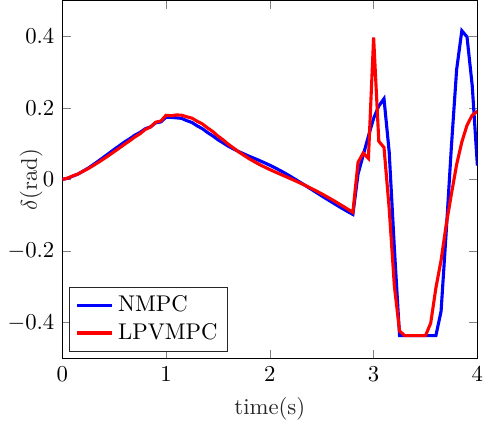}
  \caption{\centering Steering angle}
    \label{fig:SC1_obs_delta}
     \end{subfigure}
     \hfill
     \begin{subfigure}[b]{0.3\textwidth}
         \centering
  \includegraphics[scale=0.65]{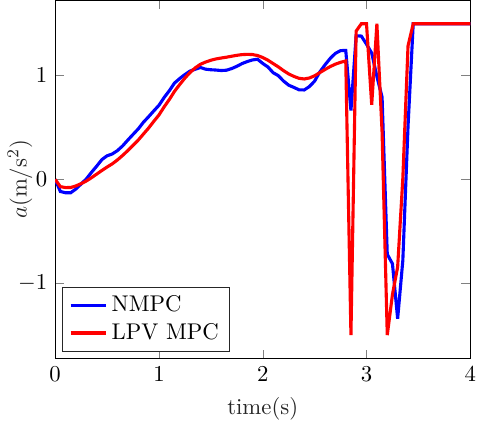}
  \caption{\centering Acceleration}
    \label{fig:Sc1_obs_a}
     \end{subfigure}
        \caption{First obstacle avoidance scenario}
        \label{fig:SC1_obs}
\end{figure*}

\begin{figure*}
     \centering
     \begin{subfigure}[b]{0.3\textwidth}
         \centering
            \includegraphics[scale=0.65]{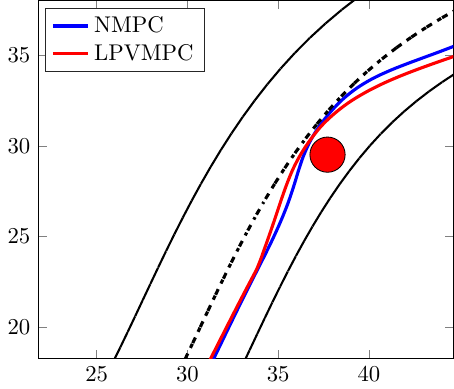}
  \caption{\centering XY position of the vehicle}
    \label{fig:SC2_obs_states}
     \end{subfigure}
     \hfill
     \begin{subfigure}[b]{0.3\textwidth}
         \centering
            \includegraphics[scale=0.65]{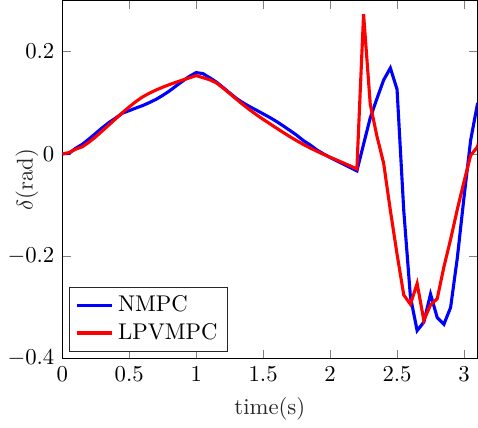}
  \caption{\centering Steering angle}
    \label{fig:SC2_obs_delta}
     \end{subfigure}
     \hfill
     \begin{subfigure}[b]{0.3\textwidth}
         \centering
 \includegraphics[scale=0.65]{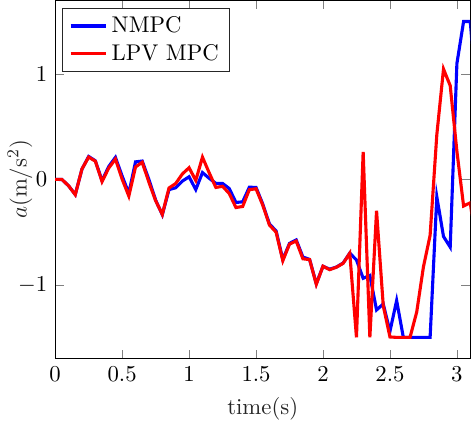}
  \caption{\centering Acceleration}
    \label{fig:SC2_obs_a}
     \end{subfigure}
        \caption{Second obstacle avoidance scenario}
        \label{fig:SC2_obs}
\end{figure*}

The presented results in~\cref{fig:SC1_obs} and~\cref{fig:SC2_obs} align with the results of the reference tracking section. Again, the results show that the LPVMPC can also perform the safe obstacle avoidance maneuver with more fluctuations in the generated inputs. Also, looking at the computation times of the LPVMPC and NMPC for obstacle avoidance scenarios, in~\cref{tab:SC1_obs} and~\cref{tab:SC2_obs} reveals that the LPVMPC is doing the task much faster than the NMPC. 


\begin{table}[h!]
    \centering
    \caption{Comparison of computation times between NMPC and LPVMPC for the obstacle avoidance scenario 1}
    \label{tab:SC1_obs}
    \setlength{\tabcolsep}{2pt}
    \begin{tabular}{ l | l | l} 
        \toprule
        \rowcolor{mydarkgray} \textbf{RT with obstacle}   & \textbf{NMPC}  & \textbf{LPVMPC}  \\ \hline
                              Average time & 0.6044  s & 0.0143  s \\
        \rowcolor{mygray}    Maximum time & 1.9183  s  & 0.0221 s \\
                             Minimum time  & 0.3157 s  &  0.0124 s   
    \end{tabular}
\end{table}

\begin{table}[h!]
    \centering
    \caption{Comparison of computation times between NMPC and LPVMPC for the obstacle avoidance scenario 2}
    \label{tab:SC2_obs}
    \setlength{\tabcolsep}{2pt}
    \begin{tabular}{ l | l | l} 
        \toprule
        \rowcolor{mydarkgray} \textbf{RT with obstacle}   & \textbf{NMPC}  & \textbf{LPVMPC}  \\ \hline
                              Average time & 0.6538  s & 0.0150  s \\
        \rowcolor{mygray}    Maximum time & 1.6906 s  & 0.0250 s \\
                             Minimum time  & 0.3422 s  &   0.0131 s   
    \end{tabular}
\end{table}

\section{Conclusion}
In this paper, an LPV model to represent a dynamic bicycle model as well as an adaptive linear realization of the obstacle avoidance constraint was suggested. 
Then, it was shown how the proposed LPV model can be used in an MPC framework to perform a reference tracking task while overtaking possible obstacles on the road. Additionally, the novel concept of scheduling trust region for smooth change of scheduling parameters in consecutive time instants in LPVMPC was proposed.

Also, the performance of the LPVMPC with trust region was compared to that of a standard LPVMPC. It was observed that the addition of scheduling trust region constraints improved the feasibility of the LPVMPC. Moreover, the proposed LPVMPC in this paper was compared to an NMPC. It was observed that the LPVMPC can perform the same task as the NMPC but with some loss in performance. On the other hand, applying the LPVMPC has the advantage that only a quadratic optimization problem needs to be solved instead of a nonlinear optimization problem, and therefore, it is a lot faster, permitting real-time controller design for such tasks that fast dynamics are evolved. 

However, this is just the beginning of looking into this problem. For example, an interesting problem that could be addressed in the future is the possibility of providing recursive feasibility for the LPVMPC or generalizing the application of the approach to avoiding dynamic obstacles, which are under our investigation. 


\bibliographystyle{unsrt}
\bibliography{autosam}

\begin{thebibliography}{10}

\bibitem{kensbock2023scenario}
Robin Kensbock, Maryam Nezami, and Georg Schildbach.
\newblock {Scenario-Based Decision-Making, Planning and Control for Interaction-Aware Autonomous Driving on Highways}.
\newblock In {\em 2023 IEEE Intelligent Vehicles Symposium (IV)}, pages 1--6. IEEE, 2023.

\bibitem{zhang2020near}
Xiaojing Zhang, Monimoy Bujarbaruah, and Francesco Borrelli.
\newblock {Near-optimal rapid MPC using neural networks: A primal-dual policy learning framework}.
\newblock {\em IEEE Transactions on Control Systems Technology}, 29(5):2102--2114, 2020.

\bibitem{kim2021backup}
Hunmin Kim, Hyungjin Yoon, Wenbin Wan, Naira Hovakimyan, Lui Sha, and Petros Voulgaris.
\newblock {Backup plan constrained model predictive control}.
\newblock In {\em 2021 60th IEEE Conference on Decision and Control (CDC)}, pages 289--294. IEEE, 2021.

\bibitem{shao2021model}
Pengyuan Shao, Yanfei Dong, and Songhui Ma.
\newblock {Model Predictive Control based on Linear Parameter-Varying Model for a Bio-inspired Morphing Wing UAV}.
\newblock In {\em 2021 International Conference on Robotics and Control Engineering}, pages 59--63, 2021.

\bibitem{brudigam2022safe}
Tim Br{\"u}digam, Robert Jacumet, Dirk Wollherr, and Marion Leibold.
\newblock {Safe stochastic model predictive control}.
\newblock In {\em 2022 IEEE 61st Conference on Decision and Control (CDC)}, pages 1796--1802. IEEE, 2022.

\bibitem{Nezami2021}
Maryam Nezami, Georg Männel, Hossam~Seddik Abbas, and Georg Schildbach.
\newblock {A Safe Control Architecture Based on a Model Predictive Control Supervisor for Autonomous Driving}.
\newblock In {\em 2021 European Control Conference ({ECC})}. {IEEE}, June 2021.

\bibitem{li2020robust}
Shuo Li and Osbert Bastani.
\newblock {Robust model predictive shielding for safe reinforcement learning with stochastic dynamics}.
\newblock In {\em 2020 IEEE International Conference on Robotics and Automation (ICRA)}, pages 7166--7172. IEEE, 2020.

\bibitem{wabersich2018linear}
Kim~P Wabersich and Melanie~N Zeilinger.
\newblock {Linear model predictive safety certification for learning-based control}.
\newblock In {\em 2018 IEEE Conference on Decision and Control (CDC)}, pages 7130--7135. IEEE, 2018.

\bibitem{Nezami2022}
Maryam Nezami, Ngoc~Thinh Nguyen, Georg Männel, Hossam~Seddik Abbas, and Georg Schildbach.
\newblock {A Safe Control Architecture Based on Robust Model Predictive Control for Autonomous Driving}.
\newblock In {\em 2022 American Control Conference (ACC)}, June 2022.

\bibitem{allamaa2022real}
Jean~Pierre Allamaa, Petr Listov, Herman Van~der Auweraer, Colin Jones, and Tong~Duy Son.
\newblock {Real-time nonlinear mpc strategy with full vehicle validation for autonomous driving}.
\newblock In {\em 2022 American Control Conference (ACC)}, pages 1982--1987. IEEE, 2022.

\bibitem{arrigoni2022mpc}
Stefano Arrigoni, Francesco Braghin, and Federico Cheli.
\newblock {MPC trajectory planner for autonomous driving solved by genetic algorithm technique}.
\newblock {\em Vehicle system dynamics}, 60(12):4118--4143, 2022.

\bibitem{mi2023integration}
Yilin Mi, Ke~Shao, Yu~Liu, Xueqian Wang, and Feng Xu.
\newblock Integration of motion planning and control for high-performance automated vehicles using tube-based nonlinear mpc.
\newblock {\em IEEE Transactions on Intelligent Vehicles}, 2023.

\bibitem{incremona2022design}
Gian~Paolo Incremona and Philipp Polterauer.
\newblock Design of a switching nonlinear mpc for emission aware ecodriving.
\newblock {\em IEEE Transactions on Intelligent Vehicles}, 8(1):469--480, 2022.

\bibitem{rodonyi2021data}
G{\'a}bor R{\"o}d{\"o}nyi, Roland T{\'o}th, D{\'a}niel Pup, {\'A}d{\'a}m Kisari, Zs~V{\'\i}gh, P{\'e}ter K{\H{o}}r{\"o}s, and J{\'o}zsef Bokor.
\newblock {Data-driven linear parameter-varying modelling of the steering dynamics of an autonomous car}.
\newblock {\em IFAC-PapersOnLine}, 54(8):20--26, 2021.

\bibitem{alcala2019lpv}
Eugenio Alcal{\'a}, Vicen{\c{c}} Puig, and Joseba Quevedo.
\newblock {LPV-MPC control for autonomous vehicles}.
\newblock {\em IFAC-PapersOnLine}, 52(28):106--113, 2019.

\bibitem{alcala2020autonomous}
Eugenio Alcal{\'a}, Vicen{\c{c}} Puig, Joseba Quevedo, and Ugo Rosolia.
\newblock {Autonomous racing using linear parameter varying-model predictive control (LPV-MPC)}.
\newblock {\em Control Engineering Practice}, 95:104270, 2020.

\bibitem{alcala2020lpv}
Eugenio Alcal{\'a}, Vicen{\c{c}} Puig, and Joseba Quevedo.
\newblock {LPV-MP planning for autonomous racing vehicles considering obstacles}.
\newblock {\em Robotics and Autonomous Systems}, 124:103392, 2020.

\bibitem{tian2022gain}
Ying Tian, Qiangqiang Yao, Peng Hang, and Shengyuan Wang.
\newblock A gain-scheduled robust controller for autonomous vehicles path tracking based on lpv system with mpc and $\text{H}_{\infty}$.
\newblock {\em IEEE Transactions on vehicular technology}, 71(9):9350--9362, 2022.

\bibitem{nezami2022robust}
Maryam Nezami, Hossam~Seddik Abbas, Ngoc~Thinh Nguyen, and Georg Schildbach.
\newblock {Robust tube-based LPV-MPC for autonomous lane keeping}.
\newblock {\em IFAC-PapersOnLine}, 55(35):103--108, 2022.

\bibitem{abbas2024linear}
Hossam~Seddik Abbas.
\newblock Linear parameter-varying model predictive control for nonlinear systems using general polytopic tubes.
\newblock {\em Automatica}, 160:111432, 2024.

\bibitem{satzger2017robust}
Clemens Satzger, Ricardo de~Castro, and Andreas Knoblach.
\newblock {Robust linear parameter varying model predictive control and its application to wheel slip control}.
\newblock {\em IFAC-PapersOnLine}, 50(1):1514--1520, 2017.

\bibitem{Rajamani2012}
Rajesh Rajamani.
\newblock {\em {Vehicle Dynamics and Control}}.
\newblock Springer {US}, 2012.

\bibitem{nezami2023design}
Maryam Nezami, Dimitrios~S Karachalios, Georg Schildbach, and Hossam~S Abbas.
\newblock {On the design of nonlinear MPC and LPVMPC for obstacle avoidance in autonomous driving}.
\newblock In {\em 2023 9th International Conference on Control, Decision and Information Technologies (CoDIT)}, pages 1--6. IEEE, 2023.

\bibitem{muraleedharan2021real}
Arun Muraleedharan, Hiroyuki Okuda, and Tatsuya Suzuki.
\newblock Real-time implementation of randomized model predictive control for autonomous driving.
\newblock {\em IEEE Transactions on Intelligent Vehicles}, 7(1):11--20, 2021.

\bibitem{de2023globally}
Oscar de~Groot, Laura Ferranti, Dariu Gavrila, and Javier Alonso-Mora.
\newblock Globally guided trajectory planning in dynamic environments.
\newblock In {\em 2023 IEEE International Conference on Robotics and Automation (ICRA)}, pages 10118--10124. IEEE, 2023.

\bibitem{MATLAB:2021}
MATLAB.
\newblock {\em {Vehicle Dynamics Blockset. Version 1.2 (R2021b)}}.
\newblock The MathWorks Inc., Natick, Massachusetts, United States, 2021.

\bibitem{MATLAB}
The~MathWorks Inc.
\newblock {MATLAB version: 9.11.0 {(R2021b)}}, 2021.

\bibitem{Lofberg2004}
J.~L{\"{o}}fberg.
\newblock {YALMIP : A Toolbox for Modeling and Optimization in MATLAB}.
\newblock In {\em In Proceedings of the CACSD Conference}, Taipei, Taiwan, 2004.

\bibitem{wachter2006implementation}
Andreas W{\"a}chter and Lorenz~T Biegler.
\newblock {On the implementation of an interior-point filter line-search algorithm for large-scale nonlinear programming}.
\newblock {\em Mathematical programming}, 106:25--57, 2006.

\end{thebibliography}
\end{document}